\documentclass[aps,prl,twocolumn,groupedaddress]{revtex4}
\usepackage{epsfig}
\usepackage{amssymb} 
\usepackage[english,polish]{babel} 

\usepackage{amsmath}

\begin{document}
\selectlanguage{english}
\title{Intrinsic Low Temperature Paramagnetism in B-DNA}
\author{S. Nakamae$^{1,2,3}$, M. Cazayous$^{1,2}$, A. Sacuto$^{1,2}$, P. Monod$^1$, H. Bouchiat$^3$}
\affiliation{$^{1}$Laboratoire de Physique du Solide (UPR 5 CNRS) ESPCI, 10 rue
Vauquelin
75231 Paris, France\\
$^{2}$ Laboratoire de Mat\'eriaux et Ph\'enom\`ene Quantiques (FDR 2437) Universit\'e Denis-Diderot 
(Paris 7) Paris, France\\
$^{3}$Laboratoire de Physique des Solides, Universit\'e Paris-Sud, B\^at 510, 91405 Orsay, France}

\date{\today}

\begin{abstract}
We present experimental study of magnetization in $\lambda$-DNA in conjunction with structural 
measurements.  The results show the surprising interplay between the molecular structures and their magnetic 
property.  In the B-DNA state, $\lambda$-DNA exhibits paramagnetic behaviour below 20 K that is non-linear in applied 
magnetic field whereas in the A-DNA state, remains diamagnetic down to 2 K.  
We propose orbital paramagnetism as the origin of the observed phenomena and discuss its relation to the 
existence of long range coherent transport in B-DNA at low temperature.
\end{abstract}
\maketitle

It is now a common knowledge that the electrical conduction in DNA is intimately linked to experimental 
factors such as molecules' base-pair sequence, type of electrodes, surrounding counter ions and number of 
water molecules \cite{Endres, Adessi, Alik1, Kelley}.  
The experimental accounts to date span a whole spectrum of conduction mechanism from insulators, 
semi-conductors, metals to proximity induced superconductors \cite{Pablo, Fink, Porath, Alik2}. 
Magnetization is an alternative, non-invasive mean to 
probe the intrinsic electronic properties of matter, 
as the measurements do not require any electrode attachments.  
Unlike the intensive experimental efforts made on electronic transport in DNA molecules, 
their magnetization has been scarcely explored due to experimental difficulties such as the 
overwhelming presence of water.  Basic questions on the intrinsic magnetic properties of DNA such as the 
magnitude of its magnetic susceptibility, $\chi_g$, have remained unclear.  
It is widely known that DNA is 
diamagnetic near room temperature with a sizable anisotropy stemming from the presence of aromatic rings of 
the base pairs whose magnitude is comparable to that of benzene \cite{Maret1,Maret2,Iizuka}.  
But how does the over-all magnetic 
state of DNA depend on intrinsic parameters (molecular structure, base-pair sequence) as well as extrinsic 
parameters such as counter ion types?  
Does DNA magnetization depend on these parameters in a way reminiscent to the electrical conduction?  
And if so, what are the consequences and implications for the usage of DNA as molecular wires?  
To answer these adjuring questions, we have studied the low temperature susceptibility and 
magnetization of randomly oriented $\lambda$-DNA molecules and its relation to their molecular structure 
(A- and B-DNA) and counter ion types (Na$^+$ and Mg$^{2+}$), 
two parameters known to greatly influence the electronic property of DNA \cite{Endres}.  
We find that the magnetization is temperature independent and diamagnetic at 
high temperatures (100 K and above) regardless of water content in both A- and B-DNA structures.
Surprisingly, once the molecules are sufficiently `wet' and thus are found in B-structure,
a paramagnetic upturn was observed at lower temperatures that is non-linear in magnetic field in addition 
to the atomic diamangetic component. 
Collectively, these observations reveal for the first time, 
the intrinsic non-diamagnetic state in DNA molecules that is intricately related to their molecular structures.

\begin{figure}[hbt]
\centering 
\epsfig{figure=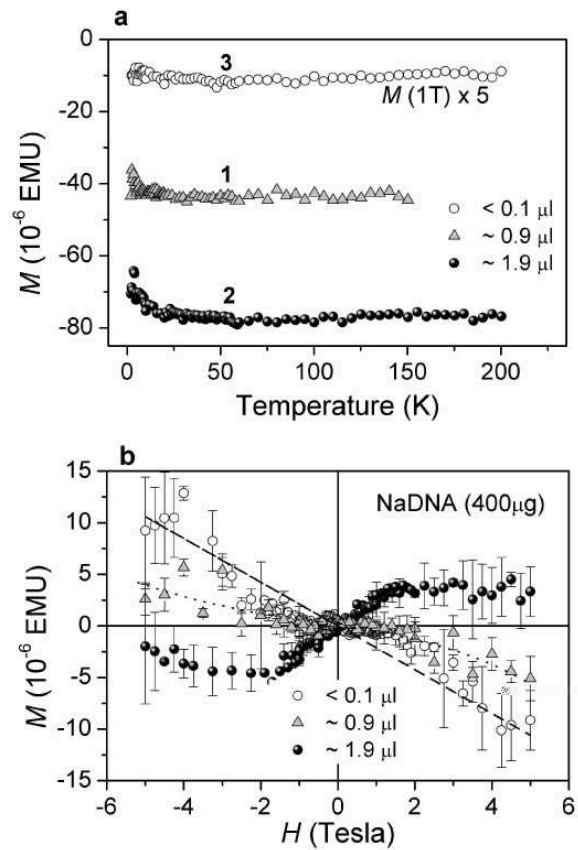, width=8.2cm}
\caption{Evolution of magnetization ($M$) of NaDNA with water content.  
(a) Magnetization of DNA and water as a function of temperature ($T$).  
For $T >$ 100 K, the change in magnetization corresponds to the temperature independent diamagnetic 
contribution from H$_2$O.  
The number in bold letters indicates the order in which the measurements were taken.  
(b) Magnetization as a function of magnetic field ($H$) measured at $T$ = 2 K without the 
contribution from water (-7.2$\times$10$^{-10}$ EMU$\cdot$ G$^{-1}\mu l^{-1}$).  The straight dashed line 
represents the diamagnetic component of NaDNA measured at $T$ = 150 K.  
The dotted curved line superimposed on the data points with 
H$_2$O = 0.9 $\mu l$ is a guide to the eye.  Error bars represent the standard deviation in the measurements.   
Some error bars on low field data, $H \leq$ 1 T, are removed in order to avoid crowding the graph.  
A Ferromagnetic component of unknown origin was detected whose magnitude saturates at $H \sim$ 7000 G 
($M \sim$ 5 $\times$  10$^{-6}$ EMU for both NaDNA and MgDNA).  
This component, however, was independent of temperature and of water content throughout the measurements 
and has also been subtracted.} 
\label{MNaDNA}
\end{figure}

\begin{figure}
\centering 
\epsfig{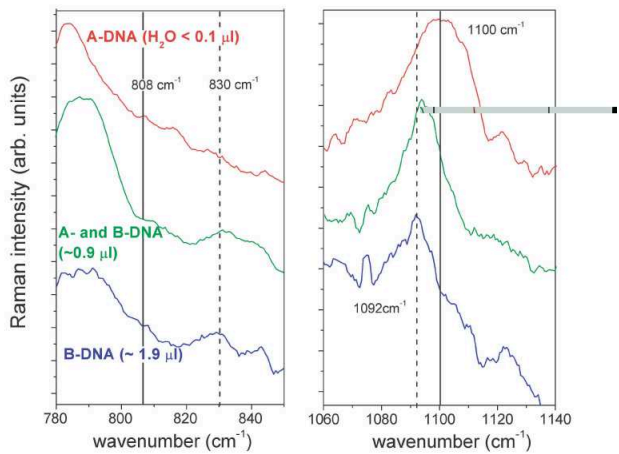}
\caption{Evolution of Raman spectra on NaDNA: 
The left panel shows the evolution of Raman band corresponding to the complex vibrational mode of the 
backbone network along the chain (5'C-O-P-O-C3').  This band shifts from 807 cm$^{-1}$ in A-DNA to 
835 $\pm$ 5 cm$^{-1}$ in B-DNA.  
The right panel shows the band associated with the symmetric stretching of the PO$_2$ moiety mode. 
This band shifts from 1100 cm$^{-1}$ in A-DNA to 1092 cm$^{-1}$ in B-DNA \cite{Deng}.}
\label{Raman}
\end{figure}

$\lambda$-DNA samples (400$\mu$g each) in two counter ion types, Na$^+$ (hereafter called NaDNA) 
and Mg$^{2+}$ (MgDNA) were prepared in quartz capillary tubes that served as sample holders for 
both magnetization (QuantumDesign MPMS-R2 SQUID magnetometer) 
and structural studies (micro-Raman spectrometer) \cite{sample}.  
$\lambda$-DNA (16$\mu$m, 48,502 base-pairs) was chosen specifically because the proximity induced 
superconductivity and the low temperature negative magnetoresistive behavior were detected previously in 
these molecules \cite{Alik1}. 
The molecular structure of DNA changes dramatically with surrounding hydration levels.  
In aqueous environment DNA molecules are in B-DNA structure where base-pairs are stacked parallel to one 
another with inter-base-pair distance of 3.2 $\rm{\AA}$ and the helix diameter of 19 $\rm{\AA}$.  
When molecules are dried, the bases become severly tilted off the helix axis and the helical diameter 
becomes broad (23 $\rm{\AA}$) \cite{Chromatin}.  
The molecular structure of the samples was transformed between the dry A-DNA and the 
wet, natural B-DNA states by adding or removing water from the samples.
In their driest states, NaDNA and MgDNA samples contained $<$0.1 and $\sim$0.3~$\mu l$ of H$_2$O, 
respectively.  
At each stage of re-hydration and/or dehydration the quartz capillaries were sealed to maintain the water 
content constant and the magnetization and the molecular structure (via micro-Raman spectroscopy) 
were studied in parallel.  Figure \ref{MNaDNA}a shows the magnetization ($M$) of NaDNA with $<$ 0.1, 
$\sim$ 0.9 and $\sim$ 1.9 $\mu l$ of water as a function of temperature ($T$) at 5 Tesla.   
Originally, the sample contained 0.9 $\mu l$ of H$_2$O.  
Then the water content was increased to 1.9 $\mu l$ and finally dried down to $\leq$ 0.1 $\mu l$.  
By subtracting the water contribution from the total magnetization at $T >$ 100 K, 
we have extracted the diamagnetic susceptibility of DNA, 
$\rm{\chi_{DNA}}$ = -0.63 $\pm$ 0.1 $\times$ 10$^{-6}$ EMU$\cdot$G$^{-1}$g$^{-1}$.  
Within the experimental accuracy, 
the diamagnetic susceptibility was found to be independent of water content, that is, 
$\chi_{A-DNA}$ = $\chi_{B-DNA}$.  
This value, determined from two NaDNA samples, is in fair agreement with the calculated atomic diamagnetic 
susceptibility of DNA, $\sim$ -0.52 $\times$ 10$^{-6}$ EMU$\cdot$ G$^{-1}$g$^{-1}$.  
At temperatures below 20 K, the magnetization of NaDNA 
containing 1.9 and 0.9 $\mu l$ of H$_2$O indicate unexpected paramagnetic up-turn that disappeared 
once the sample was dried to H$_2$O $<$ 0.1 $\mu l$.  
Figure \ref{MNaDNA}b portraying $M$ (without H$_2$O contribution) 
as a function of magnetic field ($H$) at $T$ = 2 K clearly presents this low temperature paramagnetism.  
At helium temperature, DNA in aqueous environment exhibits a magnetization crossing-over from diamagnetic 
to paramagnetic that is non linear in magnetic field.  
The magnitude of this paramagnetic increase, $\Delta M_{para}$ = $M_{tot} - M_{dia}$, 
is comparable to that of the diamagnetic contribution of DNA.  
The corresponding Raman spectra depicting 
the structural transformation from A- to B structure in NaDNA are presented in Fig. \ref{Raman} \cite{raman}.  
Among the large and highly reliable index of Raman bands corresponding to vibrational modes of DNA geometry 
and conformations \cite{Deng},  we concentrate on two bands representing the backbone vibrations 
to identify the structural state of our samples as described in the figure caption.  
While with $<$ 0.1 $\mu l$ of H$_2$O, sample was found almost purely in A-state, 
with 0.9 $\mu l$, B-DNA as well as a small signature of A-DNA were 
observed.  With further addition of H$_2$O, molecules were found entirely in the B-state.  By comparing the 
molecular structure and the magnetization of NaDNA, it appears that B-DNA is a prerequisite condition for 
the low temperature paramagnetism in DNA.  
It needs to be noted, however, that once the relative humidity, $RH$ 
(the weight of H$_2$O divided by that of dry DNA) exceeds 0.9, 
the DNA molecules assume B-DNA state \cite{Lindsay}.  
$RH$ = 0.9 corresponds to 0.36 $\mu l$ of H$_2$O in our sample, 
far less than the nominal amount of 0.9 $\mu l$ used here.  
This observation indicates that H$_2$O is not diffused uniformly due to the sample geometry and the 
preparation method.  The more rigorous investigation on the water content and the structural analysis 
on NaDNA samples via X-ray diffraction will be reported elsewhere.

\begin{figure}
\centering 
\epsfig{figure=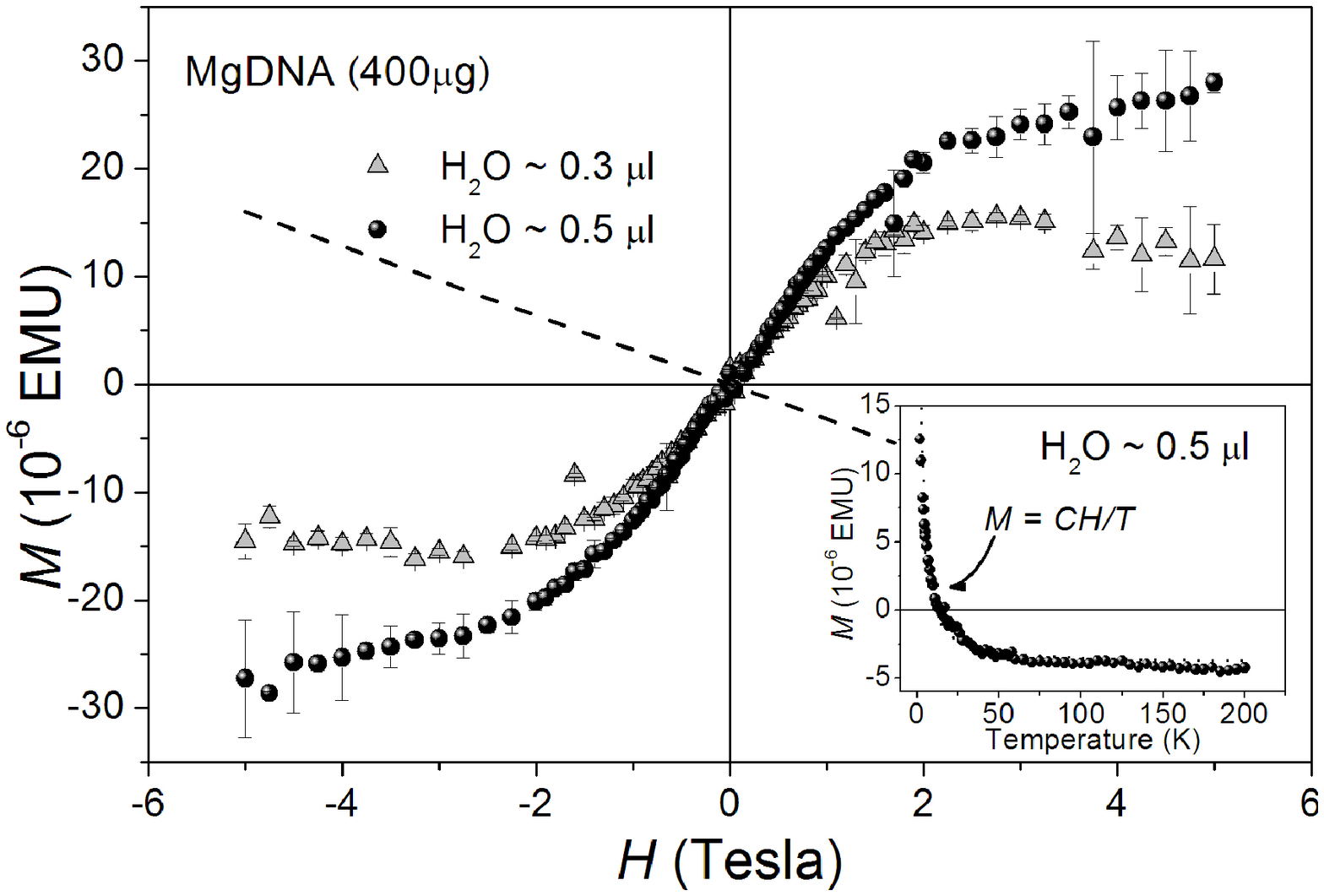, width=\linewidth}
\caption{$M$ vs. $H$ of MgDNA sample measured at $T$ = 2 K.  
 The data presented here are treated in the same manner as in Fig. \ref{MNaDNA}b.  
 Inset: $M$ as a function of temperature of wet MgDNA at $H$= 1T.  
 Dotted line represents the fit to the Curie law, $M$($T$) = $C \times H/T$, 
 with $C$ (the Curie constant) = 4.3 $\pm$  0.2 $\times$ 10$^{-9}$ J$\cdot$ K/T$^2$ 
 corresponding to $\sim$ 10$^{15}$ spins of $s$ = 1/2.}
\label{MMgDNA}
\end{figure}

In MgDNA sample, we were unable to remove H$_2$O sufficiently to create predominantly A-DNA state.  
In fact, the Raman spectra (not shown) of the MgDNA in its driest state (0.3 $\mu l$) indicated mainly B-DNA 
bands, and with $\sim$0.5 $\mu l$ of H$_2$O the molecules were found to be in purely B-DNA state.  
This is in marked contrast with NaDNA where the presence of A- and B-DNA were both detected at much higher 
water content.  
This observation is consistent with a known property of Mg$^{2+}$, {\it i.e.}, 
that prevents the transition from B- to A-DNA more efficiently than Na$^+$ ions \cite{Schultz}.  
The magnetization measurements on MgDNA with 
$\sim$ 0.5 $\mu l$ of H$_2$O preceded the measurements with 0.3 $\mu l$.  
As can be seen from Fig. \ref{MMgDNA}, a purely diamagnetic 
behaviour at low temperatures was never achieved in MgDNA in line with the observation in NaDNA.  
In the driest state, only a slight decrease in $\Delta M_{para}$ was detected.  
Furthermore, the paramagnetic magnetization was found to become independent of water content 
for H$_2$O values higher than 0.5 $\mu l$ (measured up to 2.2 $\mu l$).  
Temperature dependence of the magnetization of wet MgDNA follows the Curie law as 
shown in the inset.  Susceptibility at higher temperatures was determined to be 
-0.8 $\pm$ 0.1 $\times$ 10$^{-6}$ EMU$\cdot$ G$^{-1}$g$^{-1}$, 
larger than the value found for NaDNA sample.  
This difference ($\sim$ 0.2 $\times$ 10$^{-6}$ EMU$\cdot$ G$^{-1}$g$^{-1}$) corresponds to the diamagnetic 
susceptibility of the residual buffer ions (MgCl$_2$ and NH$_4$-Acetate).  
It is also noteworthy that the low temperature $\Delta M_{para}$ was found to be $\sim$4 times higher 
in MgDNA than in NaDNA. 

The apparition of paramagnetism in B-DNA at low temperature that is non-linear in applied field is robust.  
There are two possible origins for the observed non-linear paramagnetism: electron spin ($s$) and orbital 
magnetism.  
Assuming the electron spins (magnetic ions or hydroxyl radicals \cite{Debije}, for example) to render 
the observed behaviour, we have fit the paramagnetic component of the magnetization, $\Delta M_{para}$, 
to the Brillouin and Curie law.  
The best fits were obtained for $s$= 1/2$\sim$3/2 with the total number of spins of $\sim$10$^{15}$ for 
NaDNA and $\sim$4$\times$10$^{15}$ for MgDNA, respectively.  
Such high concentrations of spins should be detectable by 
Electron Paramagnetic Resonance (EPR) provided that the signal line width does not exceed 300G.  
The examination via EPR \cite{EPRexp} revealed no such presence in NaDNA at room temperature.  
The MgDNA sample was examined between room temperature and 4 K.  
The only EPR absorption signal was detected at $g$= 4.28 which 
can be attributed to Fe$^{3+}$ ($s$= 5/2, $g$ = 30/7) in an asymmetric crystal field.  
The number of these spins was determined to be of the order of 10$^{12}$, 
far too small to be responsible for the observed $\Delta M_{para}$ \cite{EPR}.  
Furthermore, the magnetization of H$_2$O used to humidify the samples was examined separately 
using SQUID magnetometer.  
2 $\mu l$ of H$_2$O, was found to contain $\Delta M_{para}$(5 T, 2 K) less than 10$^{-6}$ EMU.  
Therefore these experiments as well as the disappearance of paramagnetic component in A-DNA after the 
drying process exclude free radicals and magnetic impurities in water and buffer solutions from the 
possible origins of low temperature paramagnetism.

\begin{figure}
\centering 
\epsfig{figure=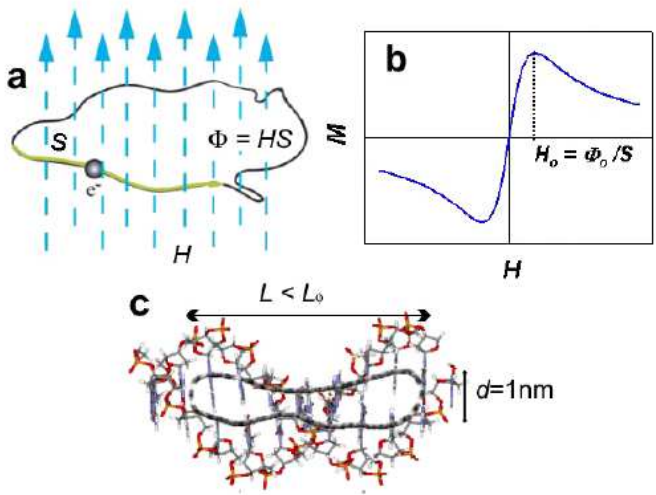, width=\linewidth}
\caption{(a) Schematic view of a mesoscopic ring flux in perpendicular magnetic field $H$.  
(b) Shape of orbital magnetization as a function of magnetic field in the non-linear 
regime.   
(c) Simplified picture of a persistent current path inside and along the helical length of a B-DNA 
molecule involving interstrand charge transfer.}
\label{orbital}
\end{figure}

Thus far, the paramagnetism appears to be an intrinsic property unique to B-DNA.  
An interesting possibility is the existence of persistent current loops along the DNA molecules on a 
mesoscopic micron scale.  The mesoscopic orbital magnetism has been shown theoretically to be 
paramagnetic and non-linear when repulsive electron-electron interactions dominate over single 
particle effects \cite{Altshuler, Ambegaokar}.   
The total magnetization of the system then follows
 %\begin{equation}
 $m(H)=\frac{\chi_L k_F \sqrt{S(T)}H}{\left[1+\left(\frac{HS(T)}{\phi_o}\right)^2\right]}$,
 %\end{equation}
where $\chi_L$ is the Landau susceptibility, and $k_F$ is the Fermi wave number.
The non-linear magnetization reaches its maximum at $H_o$ = $\Phi_o/S$($T$), 
where $\Phi_o$ = $h/e$ = 4.14 $\times$ 10$^{-7}$ G cm$^2$ is the magnetic flux quantum and $S(T)$ 
is the maximum surface area enclosed inside the coherent current loop (Fig. \ref{orbital}(a) and (b)).  
The orbital magnetism associated to the persistent currents has been already observed in mesoscopic rings 
and 2-D squares and is considered the hallmark of phase coherent transport at low temperatures \cite{Levy}. 
In our DNA samples measured at 2K, the magnitude 
of paramagnetic signal is 2 $\sim$ 4 times that of the total diamagnetic susceptibility.  
At lower temperatures the size of orbital paramagnetic susceptibility is expected to grow rapidly.  
We estimate the electron coherent length, $L$, using the experimental values from our measurements, 
$H_o$ = 1 $\sim$ 2 Tesla, via $S(T)$ = $d \times L(T)$ where $d$ is the distance between bases of B-DNA 
molecules.  
Our calculation yields the electron path on the order of 1 $\mu$m along the helical axis of the molecules.  
Such circulation of electrons enclosing a finite flux can be achieved through the combination of 
intra- and interstrand (across the hydrogen bonds) transfer of $\pi$ electrons on bases (Fig. \ref{orbital}c).  
Hydrogen bond assisted electronic exchange has already been witnessed in some organic molecules \cite{Ferrer}.  
The value of the electron path found here agrees with the coherent length of $\sim$1$\mu$m in $\lambda$-DNA 
determined by Kasumov {\it et al.} \cite{Alik1} where proximity induced superconductivity was detected 
at $T <$ 1 K.  
Our observation may also imply a coherent electron transport along the helical length of the molecule at 
low temperatures, but exclusively in B-DNA, consistent with experimental reports on the DNA electronic 
conductivity that showed higher conductivity in wet-DNA molecules \cite{Tran, Otsuka}.
Lastly, from the enhanced size of the low temperature paramagnetic signal as well as the persistence 
of B-DNA in MgDNA, Mg$^{2+}$ appears to facilitate the electron transfer inside DNA molecules.  

In summary, we have found a low temperature, non-linear paramagnetic behaviour in the B-state of 
$\lambda$-DNA molecules.  This effect is found in both DNA samples prepared with Na$^+$ and Mg$^{2+}$ 
counter ions.  The paramagnetic susceptibility of molecules prepared with Mg$^{2+}$ ions is found larger 
by a factor of 4 compared to the Na$^+$ counterpart.  The present results can be interpreted by the existence 
of a mesoscopic orbital paramagnetism in B-DNA molecules that may be related to the proximity induced 
superconductivity observed in these molecules.  Magnetization of other types and aligned DNA molecules 
should be examined in order to confirm the orbital origins of this paramagnetism. 

We thank F. Livolant, A. Bertin, D. Durand, S. Gu{\'e}ron, J.F. Allemand, D. Bensimon and V. Croquette 
for stimulating discussions and experimental guidances.


\begin{thebibliography}{00}

\bibitem{Endres} R. G. Endres {\it et al.}, Rev. Mod. Phys. {\bf 76}, 195 (2002).

\bibitem{Adessi} Ch. Adessi {\it et al.}, Comp. Nanosci. Nanotech. {\bf 2002}, 56, (2002).

\bibitem{Alik1} A. Yu. Kasumov {\it et al.}, Appl. Phys. Lett. {\bf 84}, 1007 (2004).

\bibitem{Kelley} S. O. Kelley and J. K. Barton, Science {\bf 283}, 375 (1999).  

\bibitem{Pablo} P. J. de Pablo {\it et al.}, Phys. Rev. Lett {\bf 85}, 4992 (2000).

\bibitem{Fink} H. W. Fink, C. Sch$\ddot o$nenberger, Nature {\bf 398}, 407 (1999).

\bibitem{Porath} T. Porath {\it et al.}, Nature, {\bf 403}, 635 (2000).

\bibitem{Alik2} A. Kasumov {\it et al.}, Science {\bf 291}, 280 (2001).

\bibitem{Maret1} G. Maret {\it et al.}, Biopolymers {\bf 22}, 2727, (1983).

\bibitem{Maret2} G. Maret {\it et al.}, Phys. Rev. Lett. {\bf 35}, 397 (1975).

\bibitem{Iizuka} E. Iizuka and Y. Kondo, Mol. Cryst. Liq. Cyrst. {\bf 51}, 285 (1979).

\bibitem{sample} Samples were obtained from New England Biolabs and 
Amersham Bioscience (500 $\mu$g/ml with 10mM Tris-HCl and 1mM EDTA).  
NaDNA: The original solution was diluted in 10 ml of 60/40 H$_2$O/Isopropanol solution 
containing 0.3 M of NaCl for co-precipitation.  
The solution was then centrifuged at 15 kG and at 4 $^o$C for 35 minutes.  The precipitate was rinced in 
70$\%$ Ethanol and was centrifuged again at 15 kG and at 4 $^o$C for 15 minutes.  
This procedure was repeated twice to remove the excess Na$^+$ ions.   
MgDNA: The original solution was replaced by 9 mM MgCl$_2$/20 mM NH$_4$-acetate buffer solution by dialysis.  
This solution was concentrated via centrifuging through Microcon (Millipore) filter, down to 
$\sim$ 5 mg/ml, then lyophilized for 2 hours at room temperature.  
%From UV-analysis we estimate that $\sim$ 400 $\mu$g of $\lambda$-DNA was recovered for each sample.  
The H$_2$O content of the samples were controlled by injection 
(Milli-Q distilled and deionized H$_2$O, $>$ 18 M$\Omega$) or 
by evaporation at 45 - 55 $^o$C ($\sim$ 48 hours for NaDNA and over one week for MgDNA samples).  
Quartz capillary tube sample holders were obtained from Heraeus (2 mm O.D. and 1 mm I.D.).  
The H$_2$O amount was determined using a precision scale ($\pm$ 0.1 $\mu$g).
Samples were suspended by capillary force at the mid-height of the capillary tube.  
%This configuration provided a symmetric and continuous background magnetic moment that results in 
%zero net magnetization from the sample holder.


\bibitem{Chromatin} A- and B-DNA structures can be found in, for example, 
K. E. van Holde, Chromatin, Springer Series in Molecular Biology, Springer-Verlag, 
Paris (1988).

\bibitem{Deng} H. Deng {\it et al.}, Biopolymers, {\bf 50}, 656 (1999).

\bibitem{raman} The Raman spectra were obtained with 514.5~nm excitation line of an Ar$^+$-Kr$^+$ laser 
in confocal micro-Raman configuration with $\times$10 magnification.
The scattered light was analysed using a Jobin-Yvon triple grating spectrometer (T64000) 
consisting of a holographic notch filter and liquid nitrogen cooled CCD detector. 
The radiation power at source was between 3 and 10 mW.  
Raman spectra were taken at several regions within the sample. 
Spectra shown here are the accumulated averages of 10 exposures of 30-60 seconds each.
The effective spectral resolution was less than 1 cm$^{-1}$. 

\bibitem{Lindsay} S. M. Lindsay {\it et al.}, Biopolymers {\bf 27}, 1015 (1988).

\bibitem{Schultz} J. Schultz {\it et al.}, Biophys. J. {\bf 66}, 810 (2001). 

\bibitem{Debije} M. G. Debije {\it et al.}, Radiat. Res., {\bf 154}, 163 (2000).

\bibitem{EPRexp} Bruker spectrometer equipped with an Oxford cryostat was operated at  
10GHz operation frequency, 100kHz modulation frequency and 20 G modulation amplitude.

\bibitem{EPR} The number of spins was determined by comparing the intensity of the spectra 
to that of a calibrated CuSO$_4\cdot$5H$_2$O single crystal.

\bibitem{Altshuler} B. L. Alsthuler, {\it et al.}, Phys. Rev. Lett., {\bf 66}, 88 (1991).

\bibitem{Ambegaokar} V. Ambegaokar and U. Eckern, Phys. Rev. Lett., {\bf 65}, 381 (1990).

\bibitem{Levy} L. P. L{\'e}vy, {\it et al.}, Physica B, {\bf 189}, 204 (1993).

\bibitem{Ferrer} J. R. Ferrer {\it et al.}, Chem. Mater. {\bf 13}, 2447 (2001).

\bibitem{Tran} P. Tran {\it et al.}, Phys. Rev. Lett., {\bf 85}, 1564 (2000).

\bibitem{Otsuka} Y. Otsuka {\it et al.}, Jap. Journ. Appl. Phys., {\bf 41}, 891 (2002).
\end{thebibliography}
\end{document}